\documentclass[12pt]{article}
\usepackage{amsfonts,cite}

\def\prepno#1#2
    {\thispagestyle{empty}
    \noi    \unitlength=1mm
    	\begin{picture}(178,10)
            \put(177,15){\llap{\bf #1}}
            \put(177,10){\llap{\small\rm #2}}
        \end{picture}
          }             

\def\noi{\noindent}

\def\nqq{\hspace*{-2em}}
\def\nhq{\hspace*{-0.5em}}

\def\cm{\hspace*{1cm}}

\def\al{&\nhq}
\def\lal{&&\nqq {}}

\def\eq{Eq.\,}
\def\eqs{Eqs.\,}
\def\beq{\begin{equation}}
\def\eeq{\end{equation}}
\def\bear{\begin{eqnarray}}
\def\bearr{\begin{eqnarray} \lal}
\def\ear{\end{eqnarray}}
\def\earn{\nonumber \end{eqnarray}}

\def\yy{\\[5pt] {}}

\def\eql{\al =\al}

\def\mn{_{\mu\nu}}
\def\MN{^{\mu\nu}}

\def\d{\partial}
\def\const{{\rm const}}
\def\sign{\mathop{\rm sign}\nolimits}

\def\wh{wormhole}
\def\whs{wormholes}

\def\sph{spherically symmetric}
\def\ssph{static, spherically symmetric}
\def\asflat{asymptotically flat}

\def\oR{\overline{R}}
\def\og{\overline{g}}

\def\M{{\mathbb M}}
\def\R{{\mathbb R}}
\def\S{{\mathbb S}}
\def\ME {\mbox{$\M_{\rm E}$}}
\def\MJ {\mbox{$\M_{\rm J}$}}

\begin{document}
\prepno{gr-qc/0612032}
\begin{center}

{\large\bf No realistic wormholes from ghost-free scalar-tensor\yy
		phantom dark energy}

\bigskip
{\rm Kirill A. Bronnikov$^1$ and Alexei A. Starobinsky$^2$}

\medskip
   {$^1$\small \it Centre of Gravitation and Fundamental Metrology,
   VNIIMS, 46 Ozyornaya St., Moscow 117361, Russia; \\
   Institute of Gravitation and Cosmology, 
		PFUR, 6 Miklukho-Maklaya St., Moscow 117198, Russia  }\yy
   {$^2$\small \it Landau Institute for Theoretical Physics RAS,
   		2 Kosygina St., Moscow 119334, Russia}
\end{center}

\vspace{0.5cm}

\centerline{\bf Abstract}

{\small
\medskip\noi
It is proved that no wormholes can be formed in viable scalar-tensor models
of dark energy admitting its phantom-like ($w < -1$) behaviour in cosmology,
even in the presence of electric or magnetic fields, if the non-minimal
coupling function $f(\Phi)$ is everywhere positive and the scalar field
$\Phi$ itself is not a ghost. Some special static, spherically symmetric
wormhole solutions may exist if $f(\Phi)$ is allowed to reach zero or to
become negative, so that the effective gravitational constant becomes
negative in some region making the graviton a ghost. If $f$ remains
non-negative, such solutions require severe fine tuning and a very peculiar
kind of model. If $f < 0$ is allowed, it is argued (and confirmed by
previous investigations) that such solutions are generically unstable under
non-static perturbations, the instability appearing right near transition
surfaces to negative $f$.  }

\medskip\noi

\vspace{0.8cm}
\noi
{\bf 1.} Wormholes are hypothetic objects described by nonsingular solutions
of gravitational field equations with two large or infinite regions of
space-time connected by a throat (a tunnel). These two regions may lie
in the same universe or even in different universes. The existence of stable
static or stationary (traversable, Lorentzian) wormholes can lead to
remarkable astrophysical effects as was recently emphasized in
\cite{KNS06}, as well as to the possibility of realizing hyperspace
jumps (`null-transportation') or time machines \cite{L05}. That is why it is so
important to investigate if really existing matter can produce and support
such objects (it is well known that vacuum Einstein equations do not admit
static \wh\ solutions).

It is not difficult to construct a static wormhole without worrying
about a matter source, as was done, e.g., by Morris and Thorne \cite{MS88}.
Problems arise when trying to find an explicit, internally self-consistent
matter model whose energy-momentum tensor (EMT) has a structure required
for \wh\ existence and stability. Really, it is known that such matter
should be rather exotic in the sense that its effective EMT should violate
the weak or/and null energy conditions
\beq                                                         \label{Econd}
		T_{\mu\nu} u^{\mu}u^{\nu}\ge 0
\eeq
where $u^{\mu}$ is a time-like or null 4-vector
($u_{\mu}u^{\mu}\ge 0$).\footnote
	{Our conventions are: the metric signature $(+{}-{}-{}-)$; the
	curvature tensor
	$R^{\sigma}{}_{\mu\rho\nu} = \d_\nu\Gamma^{\sigma}_{\mu\rho}-\ldots,\
	R\mn = R^{\sigma}{}_{\mu\sigma\nu}$, so that the Ricci scalar
	$R > 0$ for de Sitter space-time and the matter-dominated
	cosmological epoch; the system of units $8\pi G = c = 1$.}
Nevertheless, beginning with the 1970s \cite{br73,ell73}, a number of static
wormhole solutions supported by scalar fields in the Einstein and
scalar-tensor theories of gravity have been constructed, see, e.g.,
\cite{KNS06,L05,LL03} for references to later work.

However, in all these solutions, to violate the above energy conditions,
such models always contain at least one ghost\footnote
	{The words ``ghost'' and ``phantom'' are often used on equal
         footing in papers on gravitation and cosmology; however, for
	 clarity, we here distinguish between ``phantom'' dark energy as
	 matter with $w < -1$ (see below) and ``ghosts'' as fields with
	 negative kinetic energy.},
i.e., a classical field with a negative kinetic term, whose energy density
may become arbitrarily negative for high frequency oscillations. From the
quantum field theory point of view, this property is bad: it leads to the
dramatic possibility of generating an unlimited amount of positive energy in
the form of equal amounts of all known particles and antiparticles in {\em
laboratory,\/} accompanied by production of equal negative energy of the
ghost field in the form of particles and antiparticles of that field, too
(see \cite{CJM04} for the most recent consideration). Note that this process
requires gravitational creation of four particles from vacuum. That is why
this instability is not seen in the behaviour of linear quantum
perturbations in a classical background. Since nothing of this kind is
observed, it seems that nature somehow avoids ghosts. Moreover, serious
problems with ghosts appear even at the classical level. First, different
instabilities arise at boundary surfaces dividing ghost and normal field
behaviour which generically transform these surfaces into singular ones
\cite{St81,BG04,V05,CK06}. Second, as was recently argued in \cite{GPRS06},
cosmological models with a ghost field cannot explain the
observed large-scale homogeneity and isotropy of the Universe. Thus, even
though there exist some counter-arguments in favour of ghost fields (see,
e.g., \cite{BF06}), it is reasonable to try to avoid such fields in
modelling real or hypothetic phenomena.

The problem of phantom matter has got a new twist after the recent
discovery of the Universe late-time acceleration \cite{RP98}. A new
form of matter dubbed dark energy (DE) is needed to support this
acceleration if the Einsteinian form of gravitational field equations
is assumed (see \cite{DErev,SS06} for reviews). Moreover, it appeared
that though DE is well described by a cosmological constant in the first
approximation, some observational data, in particular, the 'Gold' supernova
sample \cite{R04}, slightly favour a phantom behaviour of DE. Namely,
the DE equation of state  $w \equiv p_{\rm DE}/\rho_{\rm DE}$ may be less
than $-1$ for small redshifts $z < 0.3$ along with crossing of the phantom
divide $w=-1$ at larger $z$ \cite{phantom04} (see \cite{SS06} for a list of
further references). Other supernova samples like the SNLS one \cite{A06} as
well as the WMAP3 data \cite{SBD06} have a cosmological constant as the best
fit but still do not exclude recent phantom DE behaviour, see
\cite{phantom06}.

Does it mean that if the transient phantom behaviour of DE will be confirmed
by future, more precise observational data, we have to introduce ghosts? No,
not at all. It is known that there do exist models without ghosts admitting
a phantom DE behaviour and even a super-accelerated expansion of the present
Universe, $\dot H > 0$, where $H(t)$ is the Hubble parameter, $H\equiv d\ln
a(t)/dt$, $a(t)$ is the scale factor of a Friedmann-Robertson-Walker
cosmological model. The simplest of them is generic scalar-tensor gravity
generalizing the original Brans-Dicke theory to the case of a non-zero
scalar potential, see \eq (\ref{Lagr}) below. It was explicitly shown in
\cite{BEPS00,GPRS06} (see also \cite{P05}) that this DE model has sufficient
freedom to describe all possible observational data on the luminosity
distance and the inhomogeneity growth factor including the possible present
phantom behaviour and smooth crossing of the phantom divide in the recent
past. More complicated models of phantom DE without ghosts can exist, too,
but they are less investigated.

Now, it becomes very important to investigate if this physically reasonable
and potentially existing kind of exotic matter may support static and stable
wormholes. This problem is solved in this paper. Following and extending
our previous considerations in \cite{vac5,BG05}, we will also add an
electromagnetic field to the scalar-tensor DE since this might be important
both for stabilizing a wormhole and for potential astrophysical applications
\cite{KNS06}. Thus, the Lagrangian density of a general scalar-tensor theory
(STT) in a Jordan-frame manifold \MJ\ with the metric $g\mn$ is taken as
\beq
	2L= f(\Phi) R + h(\Phi)g^{\mu\nu}\Phi_{,\mu} \Phi_{,\nu}
				- 2U(\Phi) - F\MN F\mn,       \label{Lagr}
\eeq
where $R$ is the Ricci scalar, $F\mn$ is the electromagnetic field tensor,
$f$, $h$ and $U$ are arbitrary functions. It is assumed that a fermion
matter Lagrangian is not coupled to $\Phi$, so that the Jordan frame is the
physical one (in particular, fermion masses are constant and atomic clocks
measure the proper time $t$ in it). We will still use the Einstein frame,
defined as a manifold \ME\ with the metric
\beq
	\og\mn = |f(\Phi)| g\mn,                             \label{conf}
\eeq
as a convenient tool for studying the properties of $g\mn$, employing
results obtainable from the corresponding Lagrangian
\beq
        2L_E = (\sign f) \bigl[\oR
                    + (\sign l) \og\MN \phi_{\mu}\phi_{,\nu}\bigr]
		    	    - 2V(\phi) - F\MN F\mn, 		\label{LE}
\eeq
where bars mark quantities obtained from or with $\og\mn$, indices are
raised and lowered with $\og\mn$ and the following relations hold:
\beq
    	l(\Phi) := fh + \frac{3}{2}\biggl(\frac{df}{d\Phi}\biggr)^2,
    \cm
    	\frac{d\phi}{d\Phi} = \frac{\sqrt{|l(\Phi)|}}{f(\Phi)},
    \cm
	V(\phi) = |f|^{-2} U(\Phi).                          \label{trans}
\eeq

The conditions of (quantum) stability and absence of ghosts in the theory
(\ref{Lagr}) are $f(\Phi) > 0$ (the graviton is not a ghost) and $l(\Phi)
> 0$ (the $\Phi$ field is not a ghost).

\medskip\noi
{\bf 2.} Let us first assume that $f(\Phi)$ and $l(\phi)$ are smooth and
    positive everywhere, including limiting points, or, equivalently,
    that $f$ and $l$ are bounded above and below by some positive constants.
    It was shown in \cite{HV97} that a static \wh\ throat
    (defined as a minimal 2-surface in a 3-manifold) necessarily implies
    violation of the null energy condition (NEC) by matter sources of the
    Einstein equations. Meanwhile, the matter sources in (\ref{LE}) always
    satisfy the NEC, hence \whs\ (and even \wh\ throats) cannot exist in
    Einstein's frame. Further, under the assumptions made, the conformal
    mapping $\og\mn = f(\Phi)g\mn$ always transfers a flat spatial infinity
    in one frame to a flat spatial infinity in another (though, the
    corresponding Schwarzschild masses may be different due to scalar field
    effects). If we suppose that there is an \asflat\ \wh\ in \MJ, then its
    each flat infinity has a counterpart in \ME, the whole manifold is
    smooth, and we obtain a \wh\ in \ME, in contradiction to the above-said.
    Thus static and \asflat\ \whs\ are absent in the Jordan frame as well.

    This simple reasoning does not even require any spatial symmetry
    assumption and means that any static \whs\ are ruled out in the theory
    (\ref{Lagr}), if everywhere $f > 0$ and $l > 0$. Moreover, no positivity
    condition or any other restriction on $U(\Phi)$ has been assumed (apart
    from that needed for the existence of asymptotic flatness).

    It should be stressed that throats are not ruled out in the Jordan frame
    since the energy conditions may be violated locally, in full analogy
    with phantom DE behaviour in cosmology. Even though the NEC holds for
    the fields $\phi$ and $F\mn$ in \ME, it can be violated for $\Phi$ in
    \MJ. Conformal mappings like (\ref{conf}) preserve the timelike or null
    character of the vectors $u^\mu$ in (\ref{Econd}), but $T\mn$ can change
    drastically. The EMT $T\mn[\phi]$ in \ME\ has its usual form;
    however, if we write the gravitational field equations in \MJ\ in the
    Einstein form, the EMT $T\mn[\Phi]$ will contain second-order
    derivatives. Nevertheless, as we see, \whs\ as global entities cannot
    appear in \MJ.

    This no-\wh\ statement can be further strengthened in several respects.
    First, the asymptotic flatness requirement may be omitted: it is
    sufficient to require the existence of two spatial infinities (which may
    be defined in terms of sequences of closed 2-surfaces with intinitely
    growing areas in the spatial sections of the space-time manifold), so
    that spatial sections of both \ME\ and \MJ\ have the topology of a
    3-cylinder $\R \times \S^2$. (Our reasoning does not cover other
    possible \wh\ geometries:  that of a ``dumbbell'', where a throat
    connects two large but finite universes and the 3-topology is $\S^3$,
    and that of a ``hanging drop'', where one of the universes is finite and
    the 3-topology is $\R^3$.)

    Second, the restriction to the static case is not necessary. Indeed,
    Hochberg and Visser \cite{HV98} extended their previous result \cite{HV97}
    on necessary NEC violation to dynamic \wh\ throats (which required a
    more general definition of a throat in terms of anti-trapped surfaces).
    In other words, even a dynamic throat cannot exist in \ME\ without NEC
    violation. We therefore can assert that even dynamic \whs\ cannot exist
    in any \MJ\ connected with such \ME\ by well-behaved (though possibly
    time-dependent) conformal factors $f$.

    One reservation should be made here: the above reasoning employs the
    fact that a regular conformal mapping transfers a spatial infinity to a
    spatial infinity and does not change the spatial topology. This is
    true under the strong conditions that we have imposed on $f(\Phi)$. It
    is known, however, that spatial topology of the same space-time manifold
    may be different in different reference frames (a well-known example is
    the appearance of de Sitter space in closed and open FRW forms).
    Speaking of dynamic \whs, we should understand that their properties
    can be drastically different in different reference frames, but we
    consider conformal mappings connecting, in a sense, similar reference
    frames in different manifolds.

    Third, the no-\wh\ statement is valid not only for STT but for any metric
    theory of gravity (e.g., high-order theories with Lagrangians containing
    $f(R)$ or $f(R, \Phi)$) whose physical manifold \MJ\ is conformally
    related to some other manifold \ME\ in which the Einstein equations hold
    with a matter source respecting the NEC, provided the conformal factor
    is everywhere smooth and positive.

\medskip\noi
{\bf 3.} Returning to STT, let us slightly weaken our assumptions and allow
    the function $f(\Phi)$ to become zero at some $\Phi = \Phi_0$. Note
    that the DE description in cosmology does not require this, but let us
    conjecture that in local configurations the scalar field $\Phi$ may reach
    values not attainable in a cosmological setting. The situation then
    becomes more complex since now the mapping (\ref{conf}) is able to
    transfer a point or surface of finite area to spatial infinity, i.e., a
    limiting surface of infinite area.

    Let us restrict ourselves to static spherical symmetry, considering the
    theory (\ref{LE}) in a space-time with the metric
\beq                                                            \label{ds}
    ds_E^2 = A(\rho) dt^2 - \frac{d\rho^2}{A(\rho)} -
    		r^2(\rho) (d\theta^2 + \sin^2\theta\, d\varphi^2)
\eeq
    and assuming $\phi=\phi(\rho)$. The Maxwell fields compatible with
    spherical symmetry are radial electric fields ($F_{01}F^{10} =
    q_e^2/r^4$) and radial magnetic fields ($F_{23}F^{23} = q_m^2/r^4$)
    where the constants $q_e$ and $q_m$ are the electric and magnetic
    charges, respectively.

    The scalar field equation and three independent
    combinations of the Einstein equations read
\bear
	 (Ar^2 \phi')' \eql   r^2 dV/d\phi,                 \label{phi}
\\
         (A'r^2)' \eql - 2r^2 V + 2q^2/r^2;                 \label{00E}
\\
              2 r''/r \eql -{\phi'}^2 ;                     \label{01E}
\\
         A (r^2)'' - r^2 A'' \eql 2 -4q^2/r^2,              \label{02E}
\ear
    where the prime denotes $d/d\rho$ and $q^2 = q_e^2  + q_m^2$. \eq
    (\ref{phi}) follows from (\ref{00E})--(\ref{02E}), which, given a
    potential $V(\phi)$, form a determined set of equations for the unknowns
    $r(\rho),\ A(\rho),\ \phi(\rho)$.

    Now, let us inquire whether or not an \asflat\ geometry of \ME\
    described by a solution to \eqs (\ref{00E})--(\ref{02E}) can be
    conformally mapped according to \eq (\ref{conf}) (where now $f \geq 0$
    is simply some function of $\rho$) to a twice \asflat\ \wh\ geometry in a
    manifold \MJ. One flat infinity in \ME\ is assumed at $\rho=\infty$,
    and it maps to a flat infinity in \MJ\ provided $f$ has a finite limit
    as $\rho \to \infty$. Another flat infinity in \MJ\ may be obtained
    either from a centre $r=0$ (where $A/r^2 \to 0$ as $r\to 0$) or from a
    horizon (where $A = 0$ at finite $r$) in \ME. A reason is that the ratio
    $A/r^2$ does not change at conformal mappings and preserves its
    geometric meaning in \MJ, where $g_{tt} = A/f$ and $-g_{\theta\theta} =
    r_J^2 = r^2/f$, and a flat infinity implies $g_{tt}\to 1$ while $r_J^2
    \to \infty$ (recall that $g\mn$ is the metric in \MJ). Thus we must have
    $A/r^2 =0$ at a preimage of a flat infinity in \MJ.

    Assuming that a centre in \ME\ is located (without loss of generality)
    at $\rho=0$ and $r(\rho) \sim \rho^a$, $a={\rm const} > 0$ at small
    $\rho$, we can evaluate the possible behaviour of $A(\rho)$ at small
    $\rho$ from \eq (\ref{02E}) and then check whether a conformal factor
    $f$ may be chosen in such a way that $\rho = 0$ is a flat infinity in
    \MJ. An inspection, performed separately for $q=0$ and $q\ne 0$,
    shows that such a choice is impossible.

    Horizons are not excluded in solutions to (\ref{00E})--(\ref{02E})
    (though the potential $V(\phi)$ must be then at least partly negative to
    conform with the well-known no-hair theorems). Moreover, in the
    scalar-vacuum case $q=0$, considering asymptotic flatness at large
    $\rho$, only one simple horizon ($A \sim \rho-\rho_h$ near the horizon
    $\rho=\rho_h$) may appear as shown in \cite{vac1}. If $q\ne 0$,
    both simple and double ($A \sim (\rho - \rho_h)^2$) horizons may appear.

    A simple horizon in \ME\ could map into a flat asymptotic in \MJ\ if
    $f \sim \rho-\rho_h$, then $r_J \sim (\rho-\rho_h)^{-1/2} \to \infty$ as
    $\rho\to \rho_h$. However, the requirement of the proper circumference
    to radius ratio at flat infinity,
\beq
	|g^{\rho\rho}| (r'_J)^2 \to 1,                      \label{flat}
\eeq
    is violated in this case: the expression in question behaves as
    $(\rho-\rho_h)^{-1} \to \infty$ instead of tending to unity.

    We conclude that {\it \ssph\ scalar-vacuum configurations in \ME\ cannot
    be conformally mapped with conformal factors $f(\rho) \geq 0$ into twice
    \asflat\ \whs.\/}

    The situation is different for double horizons. Indeed, if $A \sim
    (\rho-\rho_h)^2$, the requirement $g_{tt} = A/f \to 1$ leads to
    $f \sim (\rho-\rho_h)^2$, hence $r_J \sim (\rho-\rho_h)^{-1} \to
    \infty$, and it is straightforward to see that the expression in the
    l.h.s. of \eq (\ref{flat}) tends to a finite limit. Additional fine
    tuning (besides the special choice of parameters leading to a double
    horizon) is then required to bring this expression to precisely 1,
    otherwise there is a spatial infinity with solid angle excess or
    deficit as, e.g., in global monopole models.

    Thus, {\it some exceptional solutions to \eqs (\ref{00E})--(\ref{02E})
    with $q\ne 0$ describe metrics that can be conformally mapped into twice
    \asflat\ \wh\ metrics in \MJ.}

    Explicit examples of such solutions are yet to be found.

    The next question is: which kind of STT admits such solutions? An answer
    is easily obtained with the aid of the relations (\ref{trans}). Let us
    use the Brans-Dicke parametrization of the general STT (\ref{Lagr}),
    namely, $f(\Phi) = \Phi$, $h(\Phi) = \omega(\Phi)/\Phi$ (the Brans-Dicke
    theory as such is the special case $\omega = \const$). We also assume
    a generic behaviour of the $\phi$ field in \ME\ near $\rho=\rho_h$
    putting $\phi'(\rho_h) = f_1\ne 0$ and $A''(\rho_h) = A_2 > 0$. Then we
    find that
\beq
       l(\Phi) = \omega(\Phi) + 3/2 \approx \Phi \phi_1^2/(2A_2) \to 0
\eeq
    as $\Phi \to 0$. So, in this limit the STT approaches the boundary
    beyond which $\Phi$ would become a ghost.

\medskip\noi
{\bf 4.} Now let us further weaken our assumptions, allowing $f(\Phi)$ in
    (\ref{Lagr}) to become negative and cross zero at some point $\Phi =
    \Phi_0$. For \ssph\ solutions of the theory (\ref{Lagr}), it has been
    shown \cite{vac4} that, if $df/d\Phi \ne 0$ at $\Phi=\Phi_0$, there
    always exist such solutions that continue from positive to negative $f$.
    Moreover, such solutions (which are special with respect to the whole
    set of solutions) generically describe \wh\ geometries \cite{vac4}.
    There exist explicit examples of such continued solutions with both zero
    and nonzero charges $q$, obtained for massless conformal scalar fields
    in general relativity ($f(\Phi) = 1- (1/6)\Phi^2$, $h(\Phi)\equiv 1$,
    $U(\Phi) \equiv 0$ in (\ref{Lagr}) \cite{br73} and for more general
    non-minimally coupled scalar fields, with 1/6 replaced by an arbitrary
    coefficient $\xi > 0$ \cite{barc-vis00,BG02}. However, the stability
    studies performed by now \cite{BG02,BG04,BG05} show that such \wh\
    solutions are generically unstable under non-static monopole (\sph)
    perturbations, and the instability is related to a negative pole of the
    effective potential for perturbations situated precisely at the
    transition sphere at which $f=0$. The existence of a generic space-like
    curvature singularity at $f\to 0$ whose structure was found in
    \cite{St81} suggests that a similar instability exists for
    non-spherically-symmetric perturbations, too. Therefore, one should
    expect that for the general STT (\ref{Lagr}) transitions to negative
    values of the effective gravitational constant ($\propto 1/f$) always
    (or at least generically) lead to instabilities.

\medskip\noi
{\bf 5.} Thus, we have proved a general theorem that {\it no wormholes
    (static or dynamic) connecting two spatial infinities can be formed in
    any ghost-free scalar-tensor theory of gravity, under the condition that
    the non-minimal coupling function $f(\Phi)$ in the Lagrangian
    (\ref{Lagr}) is everywhere positive, including possible limiting
    values.}

    This result is valid in the presence of any matter whose EMT satisfies
    the null energy condition, e.g., the electromagnetic field. It is true,
    in particular, for the astrophysically relevant case of scalar-tensor
    models of dark energy admitting a phantom-like behaviour in cosmology
    ($w < -1$) \cite{BEPS00,GPRS06}. In other words, DE models of this class
    (as well as any other models conformally related to general relativity
    with everywhere positive conformal factors and without ghost fields) do
    not predict \whs.

    We have also tried to weaken the requirements and studied the possible
    behaviour of \ssph\ vacuum and electro-vacuum configurations in
    scalar-tensor gravity (\ref{Lagr}), allowing $f(\Phi)$ to reach zero or
    even become negative. It has turned out that if $f$ only reaches zero,
    twice \asflat\ \wh\ solutions in Jordan's frame can exist but only in
    exceptional cases: 1) the corresponding Einstein-frame solution must
    comprise an extreme black hole, whose double horizon is then mapped to
    the second spatial infinity in the Jordan frame, that is only possible
    with nonzero electric or magnetic fields; 2) additional fine tuning is
    necessary to avoid a solid angle deficit or excess at this second
    infinity, and 3) the theory itself should be very special: in the
    Brans-Dicke parametrization, it should hold $\omega(\Phi) + 3/2 \to 0$
    as $\Phi \to 0$. So, at the second spatial infinity, the theory
    approaches a ghost boundary and, since $f\to 0$, the effective
    gravitational constant tends to infinity. Such solutions may be of
    certain theoretical interest but can hardly be called realistic.

    Rather a wide (although still special) class of \wh\ solutions exists
    in theories where a transition to $f < 0$ is allowed. However, previous
    studies have shown that such solutions are generically unstable under
    spherically symmetric perturbations, the instability appearing due to a
    negative pole of the effective potential at the transition surface to
    $f < 0$. This pole still does not guarantee instability, and further
    studies are necessary; but even if such \whs\ can exist, their ``remote
    mouths'' are located in antigravitational regions with $f < 0$. So, they
    cannot connect different parts of our Universe but can only be bridges
    to other universes (if any) with very unusual physics.

\medskip

AS was partially supported by the Russian Foundation for Basic
Research, grant 05-02-17450, and by the Research Programme
"Astronomy" of the Russian Academy of Sciences. He also thanks the
Galileo Galilei Institute for Theoretical Physics for the
hospitality and the INFN for partial support during the period when
this project was started. KB acknowledges partial support from FAPESP
(Brazil) and RFBR grant 05-02-17478 and thanks the Institute of Physics,
University of S\~ao Paulo for kind hospitality.

\end{document}